\documentstyle[11pt,newpasp,twoside]{article}
\def\edcomment#1{\iffalse\marginpar{\raggedright\sl#1\/}\else\relax\fi}
\reversemarginpar

\begin{document}
\title{Time Dependence in Plasma Codes}
\author{S. Seager}
\affil{Institute for Advanced Study, School of Natural Sciences,
Einstein Drive, Princeton, NJ, 08540, USA}

\begin{abstract}
Time-dependent plasma codes are a natural extension
of static nonequilibrium plasma codes. Comparing relevant timescales
will determine whether or not time-dependent treatment
is necessary. In this article I outline the ingredients
for a time-dependent plasma code in a homogeneous
medium and discuss the computational
method. In the second half of
the article I describe recombination
in the early Universe as a detailed example of a problem
whose solution requires a time-dependent plasma code.
\end{abstract}

\section{Introduction}

Time-dependent plasma codes are required for any plasma where
equilibrium is not maintained due to a time-dependent physical
process. If the relevant physical process operates on a timescale $t$
that is shorter than the timescale to reach equilibrium (of atomic
ionization stages, for example) then the level populations must be
followed in a time-dependent manner.  For example, in an ionized gas
with temperatures around $10^4$~K the dominant timescale to reach
equilibrium of the gas is the hydrogen recombination timescale,
\begin{equation}
t_{rec} = \frac{1}{\alpha(T_e) n_e} = 1.15 t_4^{0.8}n_9^{-1} {\rm hours}.
\label{eq:timescale}
\end{equation}
Here $\alpha(T_e)$ is the recombination coefficient, $T_e$ is the
electron temperature, $n_e$ is the electron density, $n_9$ is the
electron density in units of $10^9$~cm$^{-3}$, and $t_4$ is the
temperature in units of $10^4$ K (Ferland 2000). There are many
examples of timescales shorter than this such as shocks or rapidly
varying radiation fields.

Some astrophysical examples where time dependent plasma codes are
useful are:

$\bullet$ the recombination epoch

$\bullet$ cooling of the first cosmological objects by H$_2$

$\bullet$ structure formation and evolution in the Universe

$\bullet$ heating of the intergalactic medium

$\bullet$ planetary nebulae and photodissociation regions

$\bullet$ the interstellar medium.

Time dependence always implies nonequilibrium level populations. In
this article I only consider the optically thin case, where
the level population's effects on the radiation field can be ignored
or treated in a simple way.

In the second part of this article I discuss recombination in the
early Universe in detail as an example of an astrophysical problem
whose solution requires a time-dependent plasma code. The time
dependence is crucial in this case because the expansion of the
Universe is much faster than the hydrogen recombination time.

\section{Time-dependent Equations}
\label{sec-maineq}
\subsection{Time-dependent Rate Equations}
\label{sec-rateeq}
The computational method in its simplest form involves solving one
basic set of equations
\begin{equation}
\label{eq:basiceq}
\frac{dn_{i,k}}{dt} = R_{populate} - R_{depopulate},
\end{equation}
constrained by particle conservation
\begin{equation}
\label{eq:pconsv}
\sum_{i} n_{i,k}(t) = N_k
\end{equation}
and charge conservation
\begin{equation}
\label{eq:cconsv}
\sum_{k} \sum_{i} i n_{i,k}(t) - n_e(t) = 0.
\end{equation}

Here $n_{i,k}$ represents the number density of atoms of species $k$
in an ionization stage $i$, $n_e$ is the number
density of electrons, and $N_{k}$ is the total number density
of all ions of a species $k$. Note that there is one redundant
equation for each $k$ and one redundant equation for each $i$ of a
given $k$.

The population and depopulation terms in equation~\ref{eq:basiceq} for
a given species $k$ and ionization stage $i$ can be described as
follows:
\begin{equation}
\label{eq:populate}
R_{populate}  =  n_{i-1,k}(\Gamma^P_{i-1,k} + 
\Gamma^C_{i-1,k} + \Gamma^{\pm}_{i-1,k})
+ n_{i+1,k}(\Gamma^{rec}_{i+1,k}
+ \Gamma^{C,rec}_{i+1,k} + \Gamma^{\pm,rec}_{i+1,k}),
\end{equation}
\begin{equation}
\label{eq:depopulate}
R_{depopulate} =
 n_{k}(\Gamma^P_{i,k}
+ \Gamma^C_{i,k} + \Gamma^{\pm}_{i,k}
+ \Gamma^{rec}_{i,k}
+ \Gamma^{C,rec}_{i,k} + \Gamma^{\pm,rec}_{i,k}).
\end{equation}

Here $\Gamma$ are the rate coefficients:
\begin{equation}
{\rm Photoionization:} \hspace{0.2in} \Gamma^P_{i,k} =
\int_{\nu_{i,k}}^{\infty}
\frac{4\pi}{h \nu}  \sigma_{i,k}(\nu) J(\nu, t)d\nu
\end{equation}
\begin{equation}
{\rm Recombination:}
\hspace{0.2in} \Gamma^{rec}_{i,k} = n_e \alpha_{i-1,k}(T_e)
\end{equation}
\begin{equation}
{\rm Collisional}\hspace{0.1in}{\rm ionization/recombination:}
\hspace{0.2in}
\Gamma^{C/C,rec}_{i,k} = n_e f^{C/C,rec}_{i,k}(T_e)
\end{equation}
\begin{eqnarray}
\label{eq:chargeexchange}
{\rm Charge-exchange}\hspace{0.1in}{\rm ionization/recombination:}\nonumber \\
\Gamma^{\pm}_{i,k} = n_{H^+} \gamma^H_{i,k}(T_e)
+ n_{He^+} \gamma^{He}_{i,k}(T_e) \\ 
\Gamma^{\pm,rec}_{i,k} = n_{H} \overline{\gamma}^H_{i,k}(T_e)
+ n_{He} \overline{\gamma}^{He}_{i,k}(T_e). 
\end{eqnarray}
Here $J(\nu, t)$ is the radiation field, $\sigma(\nu)$ is the
photoionization cross section, $\alpha(T_e)$ is the recombination
coefficient, and $h$ is Planck's constant.  The equations are coupled
among different species by the charge exchange reactions. Not shown in
the above set of equations are that the excited states $j$ can be
followed in the same framework, rather than only considering the
ground state of an ionization stage. This can be time consuming
because it involves $j$ more equations for each $i$ and $k$.  One
simplification is to only use the lower levels in the photoionization
equations and to use a pre-computed recombination coefficient that
takes into account recombination to the upper energy levels of the
atom. Another simplification is to construct several averaged atomic
energy levels (``superlevels''; see Lucy 2001) in lieu of hundreds of
excited upper levels. This works because the individual excited
states (above the first several n levels) 
usually don't dominate most rates but their collective effects
are important.  The rate equations
can also be extended to include molecules, in which case the terms
$R_{populate}$ and $R_{depopulate}$ involve reaction rates between
different molecular species (see e.g., Stancil, Lepp, \& Dalgarno
1998).


When nonequilibrium populations are involved one must take into
account their effects on the electron temperature.
If the deviations from equilibrium are strong, the temperature 
may depart from its equilibrium value. Furthermore the
time dependence is key. The radiation field is also affected
by the nonequilibrium values although in plasma codes the
radiation field is not solved for explicitly.

\subsection{Time-dependent Temperature Equation}
\label{sec-tempeqn}
In a static situation
the temperature is determined by the equilibrium between heating and
cooling processes.  In the time-dependent case the heating and cooling
processes might not have had time to reach equilibrium and so they
must be solved for together with the time-dependent rate equations. In
general one can test whether or not a given heating and cooling
process has reached equilibrium by comparing its timescale
to the time-dependent
process (shocks, variation of radiation field, wind velocity, etc.);
see \S\ref{sec:IntroRec} for an example.

The time-dependent kinetic energy equation is
\begin{equation}
\frac{d}{dt} E(t) = G(t) - L(t),
\end{equation}
where $G(t)$ are the energy sources and $L(t)$ are the energy sinks:
\begin{equation}
\label{eq:Theating}
G(t) = \sum_{k} \sum_{i} n_{i,k}(t)
\int_{\nu_{i,k}}^{\infty}
\frac{4 \pi}{h \nu} \sigma_{i,k}(\nu) J(\nu, t) h(\nu - \nu_0)d\nu,
+ G^{bb}(t)
\end{equation}
\begin{equation}
\label{eq:Tcooling}
L(t) = \sum_{k} \sum_{i} n_{i+1,k}(t) n_e(t)
\alpha_{i-1,k}(T_e)
+ L^{bb}(t) +  L^{ff}(t). 
\end{equation}
In this example the heating processes considered
are photoionization heating and
collisional line heating $G^{bb}(t)$ (the first and second
terms in equation~\ref{eq:Theating} respectively) while the cooling
processes are recombination cooling, collisional line cooling
$L^{bb}(t)$, and bremsstrahlung cooling $L^{ff}(t)$
(the first, second, and third
terms in equation~\ref{eq:Tcooling} respectively). For explicit
descriptions of these terms see, e.g. Osterbrock (1989).  While these
processes tend to be the most important heating and cooling mechanisms,
their relative importance and the importance of other processes depends on
the specific situation (i.e. $T$, density, etc.) In general it pays to
be clever by checking which heating and cooling rates
are important in advance. For example, heating and cooling
processes due to less abundant elements can usually be ignored. The
time dependence means that the temperature (=$2E/3N_{tot}k_B$) at a
previous time, together with the equations, determines the temperature
at the present time.
\subsection{The Radiation Field}
In time-dependent plasma codes the radiation field is a  key
ingredient because it governs the level populations (ionization
stages and atomic levels) and temperature. 
It is not the goal of plasma codes to
solve for the radiation field, rather it is considered a given 
parameter in an
optically thin situation.  The detailed solution of the radiative
transfer problem is a much more complex, difficult,
and different class
of problem. Still, small changes to the
radiation field can be considered.  For example, in a plasma
with a central radiation source (e.g. an HII region) 
the radiation field can be written
\begin{equation}
4 \pi J(\nu, r) = \pi F_s(\nu, R) \left(\frac{R}{r}\right)^2
{\rm exp}(-\tau(\nu, r)) + 4\pi J_d(\nu, r),
\end{equation}
where the first term on the right side is the radiation field from the 
central source modified by absorption,
and the second term is the diffuse radiation. Here $R$ is the radius
of the central star and $r$ is the radial distance from the star center.
Note that in the optically thin case the modification
to the radiation field by exponential attenuation
can easily be computed. Here the optical depth $\tau$ is simply
\begin{equation}
\tau_{\nu}(r) = \int_{R}^{r} \sum_{k} \sum_{i}
n_{i,k} \sigma_{i,k}(\nu) ds.
\end{equation}
Usually the excited levels $j$ above the ground state
do not contribute much to the optical depth.

\section{Computational Method}
\subsection{Explicit vs. Implicit Methods}
The time-dependent rate equations and the time-dependent
temperature equation form a set of coupled first-order
nonlinear differential equations. The equations are first order with
respect to the independent variable, time.  The equations are
nonlinear because of recombination and collisional terms that involve
the interaction of two species, $n_{i,k}$ and $n_e$.  The
time-dependent plasma calculation is an initial value problem where
the initial values need to be specified and the goal is to determine
the number densities and temperature
at a later time.  There are two general classes
of numerical methods used to solve initial value problems of first-order
ordinary differential equations, explicit methods and implicit
methods.

Consider the rate equations to be of the following form
\begin{equation}
\label{eq:fi}
\frac{dn_i}{dt} = f_i(t, n_1, \ldots, n_N, n_e, T),
\end{equation}
where $f_i$ is the known equation $f_i = R_{populate}-R_{depopulate}$
described in \S\ref{sec-rateeq} In the limit  
of small timesteps equation~\ref{eq:fi} can be rewritten as
\begin{equation}
\label{eq:delta}
\Delta n_i = f_i(t, n_1, \ldots, n_N, n_e, T) \Delta t.
\end{equation}
The new value $n_{new}$ can be determined with the derivative
$f_i$ to propagate the solution,
\begin{equation}
n_{new}= n_{old} + \Delta n_{i, old}.
\end{equation}
This is known as the explicit method and is also called Euler's
scheme. More sophisticated versions of it that use
derivatives at other positions, for example, 
are more useful. For example a popular algorithm is the
Runge-Kutta method.

Explicit methods can be unstable for stiff equations---those
equations where different $n_{i,k}$ are changing on very different
timescales.  Stiff equations are common in a set of rate equations
when some rates (e.g. a chemical reaction rate or an ionization rate)
are very different from other rates that control relevant number
densities. Taking a small enough timestep to satisfy the shortest
timescale is almost always impractically time consuming. Taking a
reasonable timestep can have disastrous consequences (such as values
approaching infinity or oscillating values) for species that
evolve on the timescale(s) much shorter than this reasonable
timestep. The following typical example illustrates this nicely.

Consider the equation
\begin{equation}
\label{eq:eqconsider}
\Delta n_i = -R n_i \Delta t.
\end{equation}
The new value of $n_i$ is
\begin{eqnarray}
n_{new}= n_{old} + \Delta n_{i, old} \\
n_{new} = n_{old}(1 - R \Delta t).
\end{eqnarray}
In the limit of a large timestep, $\Delta t$, $|n_{new}| \rightarrow
\infty$, an obviously unphysical value and one that certainly does
not satisfy
the constraint of equation~\ref{eq:pconsv}.

Implicit methods are commonly recommended to deal with stiff equations.
The form of the equations is the same as equation~\ref{eq:delta}
but the derivative $f$ is evaluated at the new time,
\begin{equation}
n_{new}= n_{old} + \Delta n_{i, {\bf new}}.
\end{equation}
For the above example, $n_{new}$ becomes
\begin{equation}
n_{new} = n_{old}/(1 + R \Delta t).
\end{equation}
In this case, for a large timestep $n_{new} \rightarrow 0$, the correct
solution to equation~\ref{eq:eqconsider}.  Implicit methods for a
set of nonlinear equations (as opposed to a single linear equation
like equation~\ref{eq:eqconsider}) often involve the Jacobian which
must be used in an iterative approach to solve for the set of values
$n_{new,i}$. But for the rate equations described in
equations~\ref{eq:basiceq} through \ref{eq:chargeexchange}, the
Jacobian can be computed analytically in a straightforward way.

There are many numerical methods that implement the forward or
backward Euler scheme into more sophisticated algorithms.  Many
textbooks deal with stiff equations (e.g. Press et al. 1992;
Lambert 1973) and have detailed discussions about error analysis,
adaptive stepsizes, and more.

\subsection{Numerical Considerations}
The time-dependent rate equations and the particle
and charge conservation equations mean there is one redundant
equation for each $k$ and one redundant equation for
each $i$ for a given species $k$. A practical approach is to solve
for the time-dependent rate equations (equation~\ref{eq:basiceq})
and use the particle and charge conservation equations 
(equations~\ref{eq:pconsv} and \ref{eq:cconsv}) as a check
to test the accuracy of the solution.

The input parameters for the set of rate equations
and the temperature equation are abundances,
accurate atomic data (e.g. atomic energy levels, oscillator strengths,
absorption cross sections, rate coefficients, etc.), and a good
starting solution.  A good starting solution is usually the
static solution, which is the same set of equations 
(equation~\ref{eq:basiceq} with $dn_i/dt
\equiv 0$). If there are too many equations for an algebraic
solution, the starting solution is best obtained by a Newton-Raphson
type scheme.  The output is the population number densities,
$n_{i,k}(t)$, and the electron temperature.

Algorithms with adaptive stepsize control and 
that monitor internal errors  are extremely useful.
The errors are used to ensure the solution reaches a specified
accuracy. When debugging it is useful to follow the errors (i.e.
write them to a file).  If different 
$n_{i,k}$ have the same error this indicates the problem is related to
equations that involve those two number densities
(e.g. the derivatives or terms in Jacobian).  
In general implicit methods with adaptive stepsize
control with a specified accuracy should suffice for a 
time-dependent plasma code.

\section{Recombination in the Early Universe}
\subsection{Introduction}
\label{sec:IntroRec}
Three hundred thousand years after the Big Bang the Universe became
cool enough for the ions and electrons to form neutral atoms. This was
the recombination epoch.  The Universe expanded and cooled faster than
recombination could be completed and a small but non-zero fraction of
electrons and protons remained. This is referred to the ionized
fraction at freezeout. Recombination in the early
Universe was first studied by Peebles (1968), Zel'dovich et al. (1969) and
others, and the basic framework has remained unchanged since that
time. For a complete set of references to 
both the early papers and more recent papers see Seager et al. (2000).
The recombination epoch is a good example of a problem that must be
solved by a time-dependent plasma code because the expansion timescale
of the Universe is shorter than the recombination timescale.

Figure 1 shows the difference between a
time-dependent calculation (described later in \S\ref{sec-recombeq})
and a time-independent calculation from the Saha equilibrium
equation.
\begin{figure}
\plotfiddle{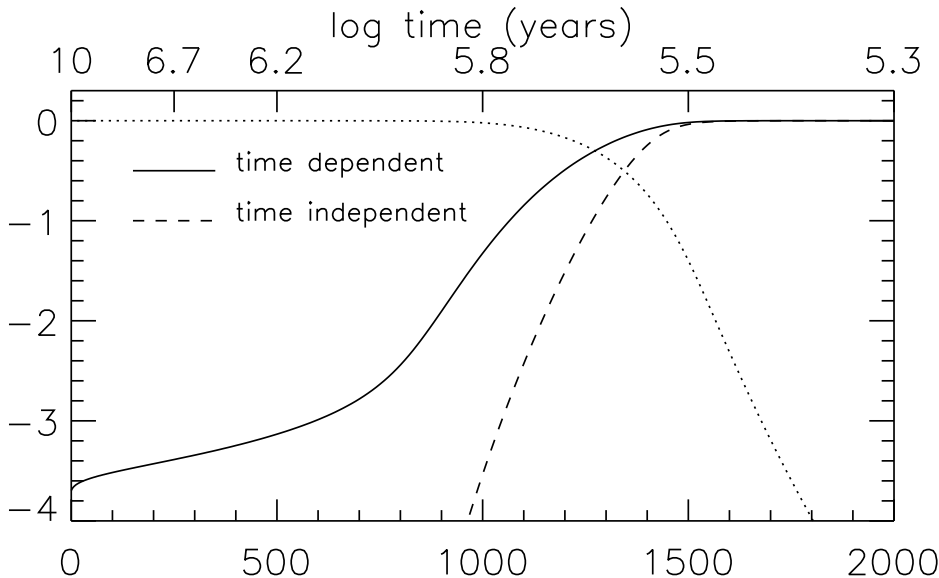}{200pt}{0}{75}{75}{-220}{-310}
\label{fig:recombbasic}	
\caption{The ionization history $x_e = n_e/n_H$ as a function of
redshift and time. The time on the upper x axis is
approximate. The solid line is $x_e$ from
the time-dependent calculation and the dashed
line is $x_e$ from the time-independent Saha equilibrium 
equation. The dotted line is the ratio of neutral H to total
H nuclei. The cosmological parameters
used in this calculation are $\Omega_m=0.25$, 
$\Lambda = 0.75$, and $\Omega_b h^2 = 0.02$.}
\end{figure}
Here $x_e = n_e/n_H$ where $n_e$ is the number density of electrons
and $n_H$ is the number of total hydrogen nuclei.  The ionization
fraction is used instead of number density because it gives a clearer
picture of recombination; expansion of the Universe changes the volume
and hence number densities and the total number density depends on the
cosmological model.  Similarly, redshift ($z$) is used instead of time
because it is independent of the cosmological model.  Redshift
is the Doppler shift of light from a receding source: $z = \Delta
\lambda/\lambda_0 = v/c = H_0 r/c$ (for low redshifts), where $v$ is the
recessional velocity, $r$ is the distance, and $H_0$ is Hubble's
constant.  Redshift and time are related by $dz/dt = -(1+z) H(z)$
where $H(z)$ is the Hubble parameter, which depends on several
cosmological parameters.  For a flat $\Omega_m = 1$ universe,
this relation is
\begin{equation}
\label{eq:redshifttime}
t = \frac{2}{3} \frac{1}{H_0 (1+z)^{3/2}},
\end{equation}
\enlargethispage{15pt} 
\hskip -0.11cm but is more complicated for other cosmologies (see, e.g., Peebles
1993). Basically redshift is a convenient parameter because it can be
measured directly and the ignorance of cosmological parameters is
hidden in usage of redshift instead of time. The radiation temperature
from adiabatic cooling is $T_R = T_0 (1+z)$, where $T_0 = 2.728$
(Fixsen et al.  1996). To illustrate the density-temperature parameter
space of recombination, Figure~2 shows the electron density as a
function of redshift and of radiation temperature for the same
ionization history and cosmological model shown in Figure~1.
\begin{figure}
\plotfiddle{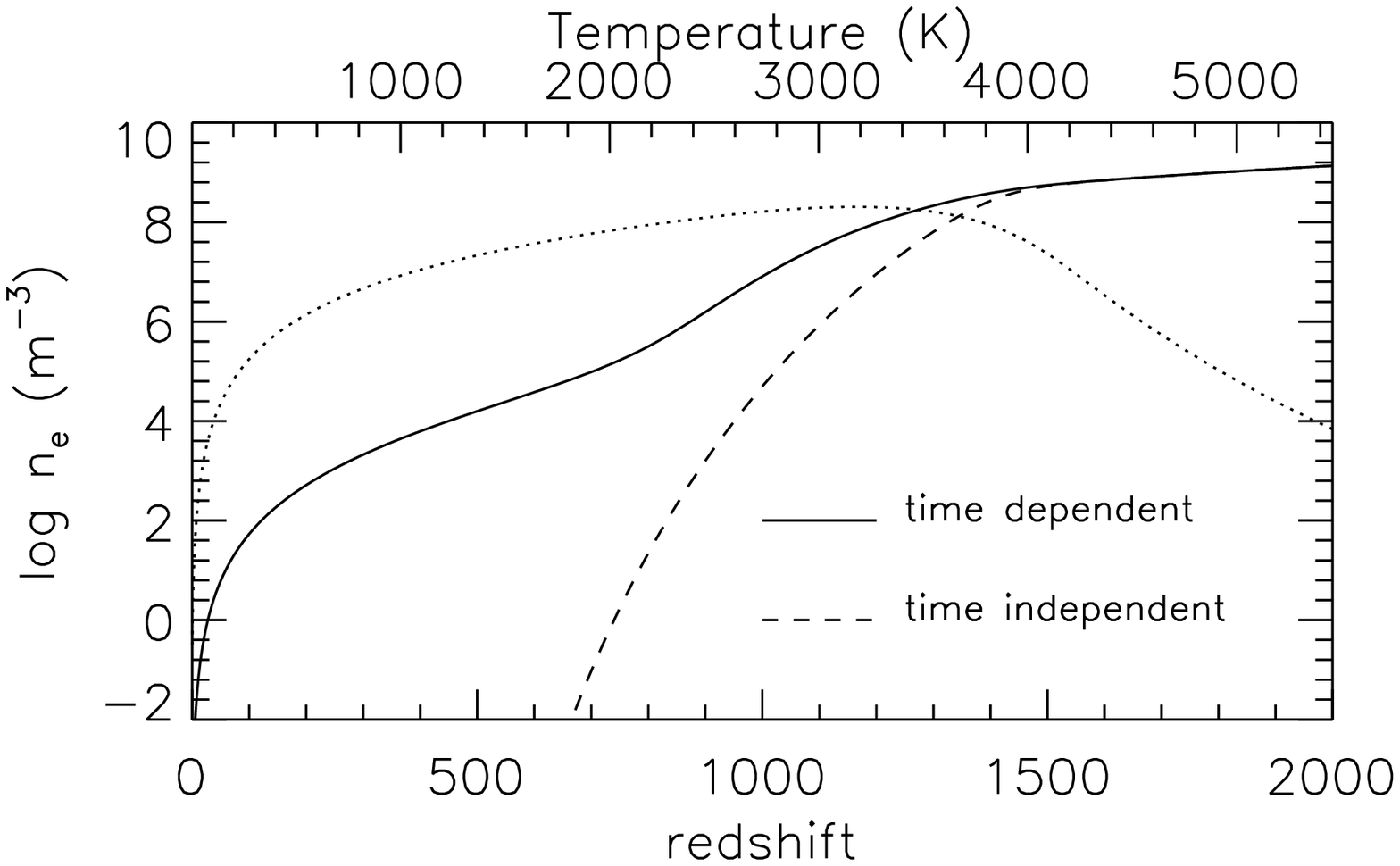}{210pt}{0}{75}{75}{-210}{-290}
\label{fig:recombnoratio}	
\caption{The same ionization history as shown
in Figure 1 but with the
number density $n_e$ as a function of redshift and of 
temperature
($T_R = 2.728 (1+z)$, and $T_M = T_R$ until $z \la 100$). The solid and dashed lines correspond
to $n_e$ from the time-dependent and time-independent calculations.
The dotted line is the number density of neutral H atoms. The
number densities decline throughout recombination because
of the expansion of the Universe.}
\plotfiddle{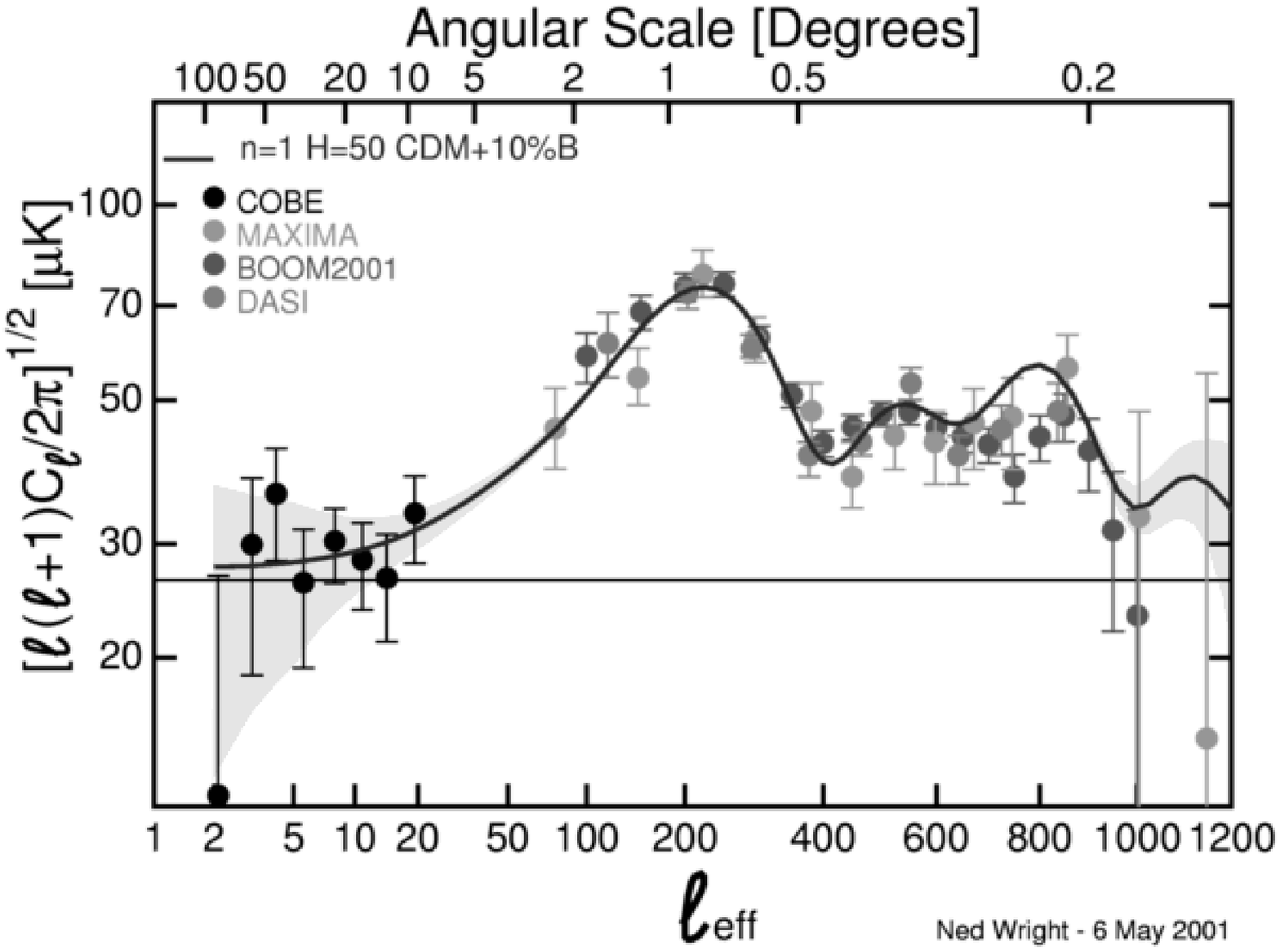}{225pt}{0}{65}{55}{-200}{-110}
\label{fig:CMB}	
\caption{Angular power spectrum of CMB anistropies from
COBE and recent experiments. Courtesy of Ned
Wright (Wright 2001).}
\end{figure}

A recent motiviation to revisit the early Universe recombination
calculation is that the ionization fraction history (shown in
Figure~1) is a basic determinant of the Cosmic Microwave Background
(CMB) spatial anisotropies which will be measured to the 1\% level
with the MAP\footnote{http://map.gsfc.nasa.gov/} satellite after its
launch in June 2001 and later with the
Planck\footnote{http://astro.estec.esa.nl/SA-general/Projects/Planck/}
satellite.  Figure~3 shows the angular power spectrum of CMB
anisotropies, where the $C_l$s are squares of the amplitudes in a
spherical harmonic decomposition of anisotropies on the sky. They
represent the power and angular scale of the CMB anisotropies by
describing the rms tempeartures at fixed angular separations averaged
over the whole sky (see e.g., White, Scott, \& Silk 1994).  These
temperature differences of the CMB at fixed angular scales on the sky
correspond to the ``seeds'' from which galaxies and other structures grew.
Cosmological parameters (such as $\Omega_0$, $\Omega_B$, $h$, etc.)
can be determined from fits (notably the relative heights and
positions of the peaks) to the power spectrum.

It is instructive to ask: when did the Universe become neutral?  As a
simple guess one might think that recombination occurred when the CMB
radiation peak had just cooled to $B_1 = 13.6$~eV, the binding energy
of the ground state of H. This corresponds to a temperature of
56,000~K, which translates to a redshift of approximately 20,000, and
a time after the Big Bang of roughly 3300 years.  A better guess would
consider that there are many, many more photons than baryons (the
ratio is $\sim$10$^9$) in the Universe.  This gives a temperature of
$\sim$7000~K or a redshift of $\sim$2500 which corresponds to a
Universe age of approximately 75,000 years.  However, due to the short
collisional timescale between H atoms (see Figure~4) the atomic
structure of H is important, and the time of recombination is
controlled by the plasma temperature.  Therefore the most reasonable
estimate comes from the Saha equilibrium equation,
\begin{equation}
\label{eq:saha}
\frac{n_{i+1} n_e}{n_i} = 
\frac{2 u_{i+1}}{u_i} 
\frac{(2 \pi m_e k_B T_M)^{3/2}}{h^3} e^{-B_1/k_B T_M},
\end{equation}
 at the point where some fraction---say 99\%---of the atoms have become
neutral. The temperature at which this occurs
is about 3000~K, which corresponds to a redshift of
approximately 1000 and a time of approximately 300,000 years
after the Big Bang. In this equation the $u_i$s are partition functions,
$m_e$ is the electron mass, $k_B$ is Boltzmann's constant,
and the other constants are as previously defined.

\enlargethispage{25pt}
It is also useful to consider different timescales that are relevant
during the recombination epoch, shown in Figure~4.
In general the relevant timescales determine whether or not a
time-dependent calculation is even necessary.  The most important
point in this example is that the expansion timescale ($t_H$) becomes
shorter than the recombination timescale early on, meaning that a
time-dependent treatment is crucial.  Also, the recombination
timescale is shorter than the Saha equilibrium timescale ($t_{Saha}$),
meaning that Saha equilibrium is not valid.  The hydrogen atom
collisional timescale ($t_{coll}$), the Coulomb timescale
($t_{Coul}$), and the proton collision timescale ($=(m_p/m_e)^{1/2}
t_{Coul}$) are all much shorter than the expansion timescale so matter
(electrons, protons, H atoms) has had plenty of time to relax to a
state of thermodynamic equilibrium. Thus we have $T_M=T_e$.  The
Compton scattering timescale ($t_{Comp}$) is faster than the expansion
timescale, meaning that the photons and electrons remain at the same
temperature throughout recombination. At $z \sim 100$ when $t_{Comp}
\gg t_H$ the matter (electrons) and radiation (photon) decouples. For
the equations that describe these timescales see Scott (1988). From
this example we see that the only equations that need to be followed
in a time-dependent manner are rate equations (to follow the time
dependence of recombination) and the matter temperature.
\begin{figure}
\plotfiddle{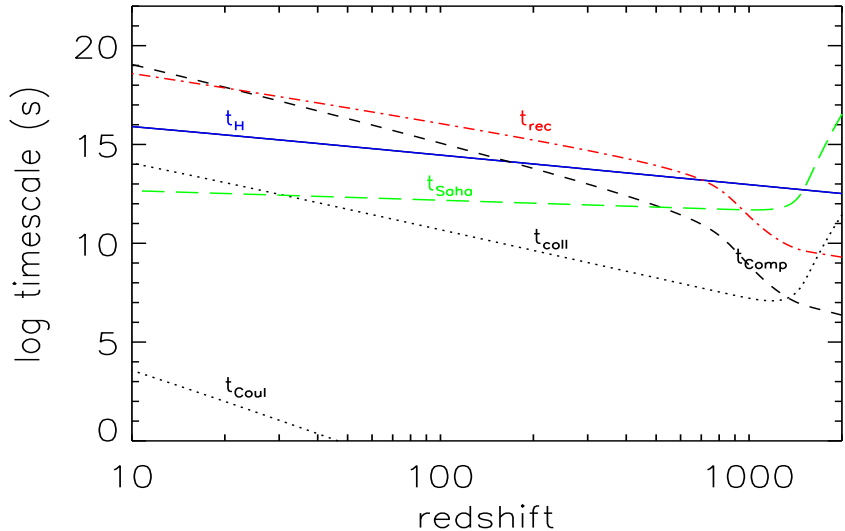}{190pt}{0}{75}{65}{-250}{-250}
\label{fig:timescales}	
\caption{Important timescales during the recombination
epoch. See text for discussion.}
\end{figure}

\subsection{The Recombination Process}
Hydrogen recombination involves protons capturing electrons and
electrons cascading down to the ground state. There are many
recombinations to and from each energy level and many bound-bound
transitions among the energy levels before there is one net
recombination to the ground state. Many people oppose the use of the
term ``recombination'' to describe the very first time in our
Universe's history that protons and electrons combined to form neutral
atoms. There were many ionizations and recombinations for every net
recombination, however, so the term recombination is still
appropriate.

The recombination process was not instantaneous (it was essentially
case B but cf., Seager et al. 2000) because of the strong but cool CMB
blackbody radiation field.  The electrons, captured into different
atomic energy levels, could not cascade instantaneously down to the
ground state. The electrons were impeded because of fast reionizations
out of excited states that were due to the huge reservoir of
low-energy photons and because of the high optical depth of the Lyman
lines and the continuum transitions to the ground state. Any Lyman
line or continuum transition to the ground state emitted a photon with
energy in which there were few blackbody photons, and this immediately
photoexcited or photoionized a neighboring atom in the ground state.
Figure~5 shows a blackbody radiation field with the energy
levels of the 13.6~eV transition and the Lyman~$\alpha$ transition
where there are few blackbody photons.  Atoms reached the
\begin{figure}
\plotfiddle{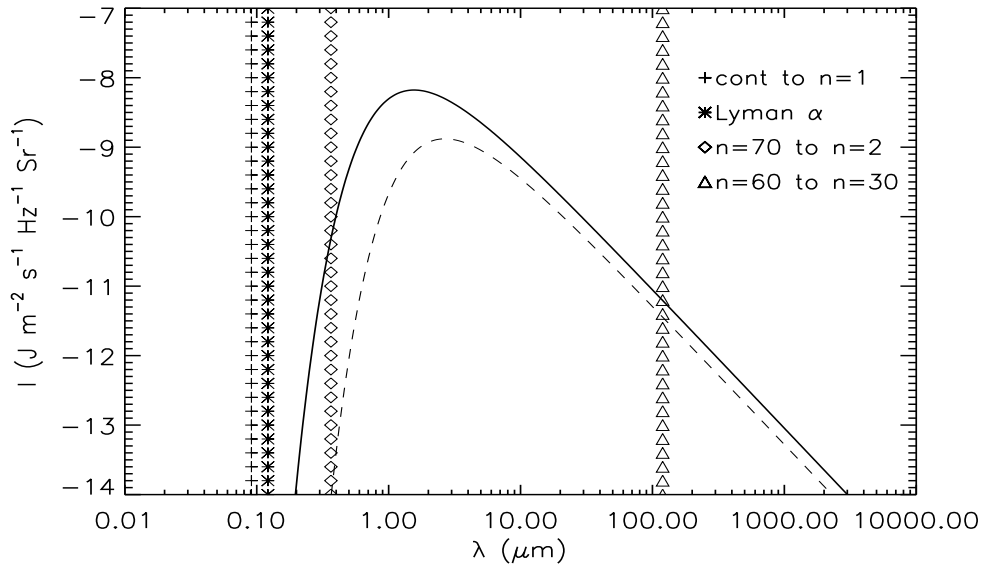}{155pt}{0}{75}{65}{-240}{-240}
\label{fig:bb}	
\caption{The solid line is the blackbody radiation intensity at
$z=1200$, which corresponds to $T_R=3274$~K. The dashed line is the
blackbody intensity at $z=500$ or $T_R=1364$. The symbols show the
wavelengths (i.e. energies) of different H atom transitions.  Here
``cont to n=1'' is the 13.6~eV continuum to ground state transition
(the binding energy of the ground state H atom).  At the wavelengths
of the continuum to n=1 transition and the Lyman alpha
transition---which are in the high-energy tail of the blackbody
radiation field---there are very few blackbody photons.  For the n=60 to
n=30 transition (as an example) there are always plenty of blackbody
photons throughout recombination and so the net rate of this
transition is always the equilibrium one---zero.  Other high-energy
transitions such as n=70 to n=2 are at a wavelength where there are
many blackbody photons early on during the recombination epoch (here
e.g. at $z=1200$) but not at later times (here e.g. at $z=500$). Thus
these types of high energy transitions go out of equilibrium at some
point during recombination and cause a faster recombination (partly
due to a faster cascade) compared to the equilibrium scenario.}
\end{figure}
\enlargethispage{20pt}
\noindent ground state either by the 2s--1s two-photon process or through the
cosmological redshifting of the Ly$\alpha$ line photons. The 
cosmological redshifting occurs because as space expands the frequency
of the photons changes---possibly enough to be redshifted out of
interaction frequency with the Lyman $\alpha$ line.  Because both of
the rates from n=2 to the ground state (the 2s--1s two-photon process
and the cosmological redshifting) were much slower than the net
recombination rate to n=2, a ``bottleneck'' occurred that slowed down
the entire recombination process.


\subsection{Recombination Equations}
\label{sec-recombeq}
Recombination in the early Universe can be computed using the
formalism described in \S\ref{sec-maineq}

\subsubsection{Time-dependent rate equations}
We are interested in the time evolution of the H n-level populations,
in addition to protons (ionized H) and electrons.  We have one time-dependent rate equation
for each atomic energy level of H,
\begin{equation}
\label{eq:recleveleqn}
(1+z)\frac{dn_{j}(z)}{dz} = 
-\frac{1}{H(z)}
\left[(n_{\rm e}(z) n_{\rm c}(z) P_{{\rm c}j} - n_j(z) P_{jc}) + 
{\rm \sum_{{\it j}=1}^{N}} \Delta R_{ji}\right]
+ 3 n_j(z),
\end{equation}
and also equations for the proton number density ($n_p$) and electron
number density (but if only H is being considered $n_p = n_e$ so only
one equation need be followed).  Here $\Delta R_{ji}$ is the net
bound-bound rate between bound states $i$ and $j$, and the $P_{j{\rm
c}}$ are the rate coefficients between bound levels $j$ and the
continuum $c$: $P_{ij} = R_{ij} + n_{\rm e} C_{ij}$ and $P_{j{\rm c}}
= R_{j{\rm c}} + n_{\rm e} C_{j{\rm c}}$, where $R$ refers to
radiative rates and $C$ to collisional rates.  Here the $n$s are
physical (as opposed to comoving) number densities: $n_j$ refers to
the number density of the $j$th excited atomic state, $n_{\rm e}$ to
the number density of electrons, and $n_{\rm c}$ to the number density
of a continuum particle (i.e. proton in this case). For convenience we
use redshift $z$ instead of time (see
equation~\ref{eq:redshifttime} and the accompanying discussion). $H(z)$ is the Hubble factor and the
extra term $3n_j(z)$ comes from the fact that space is expanding.  For
a 300-level H atom---where the number of n levels needed is determined
by a thermal broadening cutoff---there are 300 such equations and
together they involve hundreds of bound-free transitions and thousands
of bound-bound transitions.  The number of n levels that should be
considered depends on the thermal broadening cutoff.  We do not
consider individual $\ell$ states (with the exception of 2s and 2p),
but assume the $\ell$ sublevels have populations proportional to
(2$\ell$ + 1). The $\ell$ sublevels only deviate from this
distribution in extreme nonequilibrium conditions (such as planetary
nebulae).  Dell'Antonio \& Rybicki (1993) looked for such $\ell$ level
deviations for n$\,{\leq}\,10$ and found none.  For n$\,{>}\,10$, the
$\ell$ states are even less likely to differ from an equilibrium
distribution, because the energy gaps between the $\ell$ sublevels are
increasingly smaller as n increases. However the $\ell$-level time
evolution could easily be included in the plasma code but would be
consume an unreasonable amout of computer time.

The photoionization equation,
\begin{equation}
\label{eq:photoionize}
\int_{\nu_{j}}^{\infty}
 \frac{4\pi}{h \nu} \sigma_j(\nu) B(\nu,T_R)d\nu,
\end{equation}
and the recombination equation,
\begin{equation}
\left( \frac{h^2}{2 \pi m_e k_B T_M}\right)^{3/2} \frac{g_j}{2 g_k} e^{E_j/k_B T_M}
 \int_{\nu_{j}}^{\infty}\frac{4\pi}{h \nu} \sigma_{j}(\nu)
 \left(\frac{2h \nu^{3}}{c^{2}} + B(\nu,T_R)\right)
 {\rm e}^{-h \nu/k_B T_M}d\nu,
\end{equation}
are the familiar ones, but note the separate consideration of $T_M$
and $T_R$. Here $c$ is the speed of light, $g$ is the statistical
weight, $E$ is the ionization energy and the other constants and
variables have their usual meanings. Here we have replaced the usual
$J(\nu, t) = B(\nu,T_{\rm R})$ (see the below discussion on the radiation
field) because $T_R$ is a function of $z$ and
hence $t$.  Collisional rates are much smaller than radiative rates
(due to the large reservoir of photons), so while they are usually
important for a plasma of this temperature and density, they can
actually be ignored in the early Universe recombination calculation.

Every time a situation involves nonequilibrium level populations
or ionization stage populations---and
time-dependent cases are no exception---one must consider the
effects of nonequilibrium populations on the temperature
and on the radiation field.

\subsubsection{Time-dependent temperature equation}
The radiation temperature during recombination is determined
from adiabatic cooling of radiation, $T_R =
T_0 (1+z)$, where $T_0=2.728$ is today's temperature of the CMB
determined from COBE measurements (Fixsen et al. 1996).  
The time-dependent matter temperature equation for recombination in
the early Universe is
\begin{equation}
\label{eq:rectemp}
(1+z)\frac{dT_M}{dz} = \frac{8\sigma_T
U}{3H(z)m_e c}\, \frac{n_e}{n_e +
  N_H + N_{He}}\,(T_M - T_R) + 2T_M,
\end{equation}
where $\sigma_T$ is the Thomson scattering cross section and $U$ is
the energy density. The first term on the right side is Compton
cooling and the second term comes from adiabatic cooling of an ideal
gas. Other heating and cooling terms such as ionization heating,
bremsstrahlung cooling, and others mentioned in \S\ref{sec-tempeqn}
are negligible and need not be considered. In general one can
determine in advance of a calculation whether or not different heating
or cooling terms are important from comparing timescales.  For example
in this case the Compton cooling timescale becomes longer than the
expansion timescale at $z \sim 100$ (see Figure~4) and including
Compton cooling in the time-dependent matter temperature equation
keeps track of when $T_M$ and $T_R$ differ. Judging from Figure~4
it is not necessary to include the evolution of the matter temperature
for a first order calculation, since it has little effect at $z \ga
100$. Nonetheless, the $T_M$ and $T_R$ difference is significant
enough to affect the ionization fraction at freezeout by a few percent
at low $z$.

\subsubsection{The Radiation Field}
The time-dependent radiation field is 
\begin{equation}
\label{eq:radfield}
 (1+z)\frac{dJ(\nu,z)}{dz} =
 3J(\nu,z)
 -\frac{c}{H(z)}\left[j(\nu,z) - \kappa(\nu,z)J(\nu,z)\right],
\end{equation}
where $j(\nu, z)$ is the emission coefficient and $\kappa(\nu, z)$ is
the absorption coefficient.  It would be extremely difficult and time
consuming to solve this time-dependent radiative transfer equation at
each redshift for many frequencies. During the recombination epoch the
background radiation field was very smooth; it is generally a
blackbody radiation field and today the CMB is a blackbody as measured
by COBE. Furthermore the blackbody thermal spectrum is preserved in
the expansion of the Universe; the time-dependent radiation field is
simply a blackbody radiation field determined by $T_R= 2.728 (1+z)$. This means
that instead of solving equation~\ref{eq:radfield} we can just use
the blackbody intensity.  However, the extreme trapping of Lyman line
photons means there are significant distortions to the blackbody
radiation field.  In this case we can treat these distortions by the
Sobolev escape probability---this is a solution of the radiative
transfer equation in the presence of moving media, which in this case
is the expanding Universe. This approach was first used by
Dell'Antonio and Rybicki (1993) and makes the recombination in the
early Universe problem tractable.  The term $\Delta R_{ij}$ will
depend on the escape probability from Sobolev theory for transitions
where it is appropriate (the Lyman lines).

\subsubsection{Numerical Considerations}
A natural starting solution to the recombination in the early Universe
problem is at high redshift (i.e. early times and high $T$) where all the
hydrogen is ionized and the starting solution is
$n_e = n_p = N_H$ and $n_j = 0$. Because not
exactly all of the hydrogen is ionized at $z \sim 2000$ it is better
to instead use the solution of the static rate equations using a
Newton-Raphson scheme with the input to that as $n_e = n_p = N_H$ and
$n_j = 0$.

When using an integration algorithm with a specified accuracy it helps to 
not include the errors of very small level populations
in the consideration of the
size of the next timestep. I use the requirement to ignore
the errors if $n_j/N_H \la 10^{-13}$
because populations that small are not relevant
to the calculation anyway, but they are still important
enough at other times to keep their number densities
in the system
of rate equations. Otherwise the accuracy requirements
are too stringent and the integration will approach a
very tiny stepsize and a very long computational time.

\subsection{The ``Standard Recombination Calculation''}
The ``standard'' methodology forgoes the detailed numerical
computation described in \S\ref{sec-recombeq} and considers an
``effective three-level atom'' with a ground state, first excited
state (n=2), and continuum, with the n $>$ 2 states represented by a
recombination coefficient. A single ordinary differential equation can
then be derived from the rate equations shown in
equation~\ref{eq:recleveleqn} to describe the ionization fraction
(see Peebles 1968, 1993; Seager et al. 1999). Many assumptions go into
this derivation, including the following: that H excited states are in
equilibrium with the radiation; that stimulated de-excitation is
negligible for the Ly$\alpha$ transition; that a simple recombination
coefficient can be used; that every net recombination results in a
ground-state atom, so that the ground-state number density $n_1 = N_H
- n_p$; that the Ly$\alpha$ redshifting can be dealt with using a
simple escape probablility; that collisional processes are negligible;
and that He can be ignored.  Only the assumption that the upper levels
of the H atom are in equilibrium with the radiation is not valid. This
is described in the subsection below.  See Peebles (1968) or Seager et
al. (1999) for the single ordinary differential equation that
describes the ionization fraction history.

\subsection{The Exact Recombination History}
Figure 6 shows the recombination history for the ``modern'' detailed
numerical computation compared to the standard, single ordinary
differential equation calculation. The modern, more detailed
calculation has a faster recombination rate which results in a 10\%
smaller ionization fraction at freezeout compared to the standard
calculation. This in turn has an effect on the CMB power spectrum of
anisotropies of a few percent.  Note that the
cosmological parameters (and other second order effects) determine the
shape of the power spectrum, and to derive them accurately the
recombination history must be known to high accuracy.
\begin{figure}
\plotfiddle{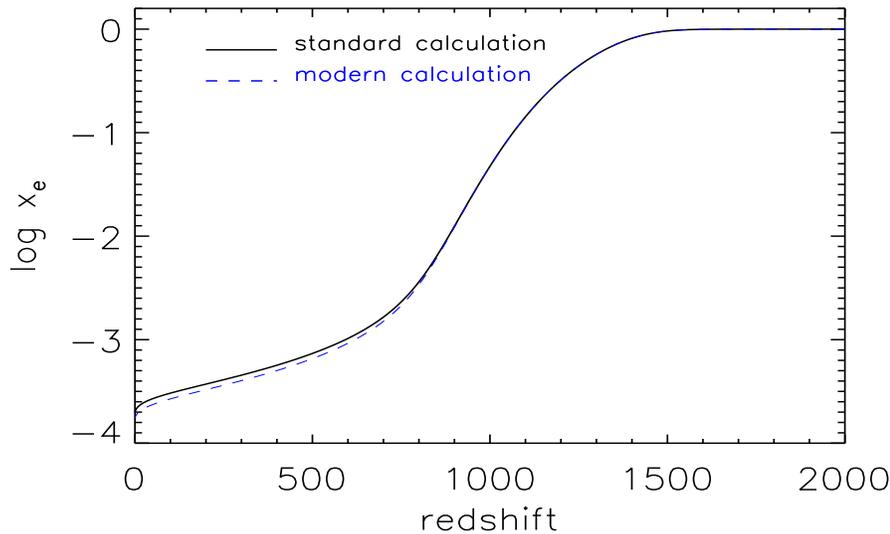}{182pt}{0}{75}{65}{-250}{-254}
\label{fig:xeoldnew}	
\caption{Ionization histories from the standard
recombination calculation compared to the more detailed
``modern'' calculation described in this article.}
\end{figure}
In the standard case that assumes equilibrium among the excited states
n $>$ 2 the net bound-bound rates are by definition zero, and this is
an implicit assumption in deriving the single ordinary differential
equation used in the standard calculation. We find that at $z \la
1000$, the net bound-bound rates become different from zero because,
at low temperatures, the cool blackbody radiation field means that
there are few photons for photoexcitation of high-energy transtions
(e.g. 70--10, 50--4, etc.). This is shown in Figure~5 where the
vertical lines show the wavelengths of various H atomic transitions,
the solid line is the blackbody intensity near the beginning of
recombination, and the dashed line is the blackbody intensity towards
the end of recombination. Because there are few available photons for
the high-energy transitions, spontaneous de-excitation dominates those
bound-bound transitions, causing a faster downward cascade to the n=2
state. In other words, once an electron is captured at, say, n=70, it
can cascade down to the n=2 state faster than in the equilibrium case
because few photons are around to photoexcite it. In addition, the
faster downward cascade rate is faster than the photoionization rate
from the upper state, and one might view this as radiative decay
stealing some of the depopulation ``flux'' from photoionization. Both
the faster downward cascade and the lower photoionization rate
contribute to the faster net recombination rate.

\subsection{Helium}
\enlargethispage{7pt}
With larger ionization potentials, HeII and HeI recombined before
H. They can be included in the same system of equations in the same
framework that has been described in this article.  These are less
important for the calculation of the CMB anisotropies, so have
generally been paid less scrutiny than H recombination. See Seager et
al. (1999) and Seager et al. (2000) for a detailed description of He
recombination.

\section{Summary}
Time dependence in plasma codes is straightforward to implement if one
already understands the set of static nonequilibrium rate equations.
The same equations that are used in static nonequilibrium plasma codes
can be used in an integration scheme to follow the time evolution of
the number densities. The temperature equation, with heating and
cooling processes should also be evolved with time if the relevant
heating and cooling processes operate on a timescale longer than the
physical process that motivates time dependence in the first place.
An implicit method or other method to treat stiff
equations---equations with number densities that are changing on very
different timescales---is usually necessary.  Numerical algorithms
with adaptive stepsize control and that monitor errors are helpful.
There can be high gain for this relatively straightforward method.

\acknowledgements
I would like to thank Gary Ferland and the conference organizing
committee for
a very interesting and useful conference.
I would also like to thank Dimitar Sasselov and Douglas Scott
for useful contributions and to
acknowledge support from the W.M. Keck Foundation.

\end{document}